\documentclass[pra, a4paper, showpacs,superscriptaddress]{revtex4-1}
\usepackage{amsmath,amscd,amsfonts,amssymb, amsbsy, color,bbm, bm}
\usepackage{enumitem}
\usepackage{natbib}

\numberwithin{equation}{section}
\usepackage{natbib}
\setcitestyle{square, comma, authoryear,sort&compress}
\begin{document}
	\title{From CCR to L{\'e}vy Processes: An Excursion in Quantum Probability}  
\author{K. R. Parthasarathy}\email{krp@isid.ac.in}
\affiliation{Indian Statistical Institute, Theoretical Statistics and Mathematics Unit,Delhi Centre,
7 S.~J.~S. Sansanwal Marg, New Delhi 110 016, India} 
\date{\today}
\begin{abstract}	 
 This is an expositary article telling a short story made from the leaves of quantum probability with the following ingredients: 
\begin{enumerate}[label=(\roman*)]
\item A special projective, unitary, irreducible and factorizable representation of the euclidean group of a Hilbert space known as the Weyl representation. 
\item The infinitesimal version of the Weyl representation includes the Heisenberg canonical commutation relations (CCR) of quantum theory. It also yields the three fundamental operator fields known as the creation, conservation and annihilation fields. 
\item The three fundamental fields, with the inclusion of time, lead to quantum stochastic integration and a calculus with an It{\^o}'s formula for products of differentials. 
\item Appropriate linear combinations of the fundamental operator processes yield all the L{\'e}vy processes of classical probability theory along with the bonus of It{\^o}'s formula for products of their differentials.  
\end{enumerate}

\vskip 1.5in 
 
\begin{center}
{\large\bf\em In memory of S.~D.~Chatterji}
\end{center}
\end{abstract}
\maketitle 
	
\section{Introduction} 
Mark Kac once remarked {\em ``Probability theory is measure theory with a soul"}. But this soul is almost lost in Bourbaki's {\em Elements of the history of mathematics}~\cite{B}, but for a discussion of brownian motion, Wiener measure, the theorems of Daniell, Kolmogorov and Prohorov on integration in non locally compact spaces. Quantum probability is a more inclusive probability theory, which draws its sustenance from classical probability as well as the statistical ideas that arise in the development of quantum mechanics and the currently fashionable theory of quantum computation and information. The spectral theory of selfadjoint operators in Hilbert space and commutation relations among operators have a profound influence in shaping this new statistics. One may recognize it in the famous book {\em Mathematical foundations of quantum mechanics} by von Neumann~\cite{vN}. It is interesting to note that its German edition appeared in 1932, one year before the publication of the German edition of Kolmogorov's {\em Foundations of the theory of probability}~\cite{K}. In the present exposition we shall not touch the subject of quantum computation and information. The interested reader may find sumptuous material on this subject in the book by Nielsen and Chaung~\cite{NC}. 

We begin with a brief discussion of the relationship between operators in Hilbert space, probability distributions and stochastic processes. All the Hilbert spaces that appear in this article are complex and separable. Let $X$ be a selfadjoint operator (bounded or unbounded) in a Hilbert space $\mathcal{H}$ with the spectral representation, 
\begin{equation}
\label{1.1}
X=\int_{\mathbb{R}} x\, P^X(dx),
\end{equation} 
 where $P^X$ is a spectral measure on the Borel $\sigma$-algebra of the real line ${\mathbb{R}}$. Suppose $u$ is a unit vector in $\mathcal{H}$. We say that $X$ is a real-valued {\em observable}, $u$ is a {\em state} and for any Borel subset $E$ of ${\mathbb{R}}$, the spectral projection $P^X(E)$  is the {\em event} that the value of the observable $X$ lies in $E$ and  the nonnegative scalar $\langle u\vert P^X(E)\vert u\rangle$ (in Dirac notation)  is the probability of the event $P^X(E)$ in the state $u$. Then one obtains the probability measure $\mu$ given by   
 \begin{equation} 
 \label{1.2}
 \mu(E)=\langle u\vert P^X(E)\vert u\rangle 
 \end{equation}
 on the Borel $\sigma$-algebra of $\mathbb{R}$. We say that $\mu$ is the {\em distribution} of the observable $X$ in the state $u$ and the scalar $\langle X\rangle$ defined by 
 \begin{eqnarray*}
 \langle X\rangle = \int_{\mathbb{R}}\, x\, \mu(dx)=\langle u\vert X\vert u\rangle 
 \end{eqnarray*} 
 is called the {\em expectation} of $X$, whenever it is finite. If $\phi$ is a real or complex-valued Borel function, $\phi(X)$ is a real or complex-valued observable with expectation 
 \begin{eqnarray*}
 \langle \phi(X)\rangle =\int_{\mathbb{R}}\, \phi(x)\, \mu(dx)=\langle u\vert \phi(X)\vert u\rangle 
 \end{eqnarray*} 
 if it exists. In particular, the characteristic function of $X$ in the state $u$ is given by 
 \begin{equation}
 \label{1.3}
 \langle e^{itX}\rangle= \int_{\mathbb{R}}\, e^{itx}\, \mu(dx)=\langle u\vert e^{itX}\vert u\rangle
 \end{equation} 
 Equations (\ref{1.1})-(\ref{1.3}) show that an observable $X$ can be viewed as a selfadjoint operator, or a spectral measure on $\mathbb{R}$ or as the one parameter unitarty group, each having its own statistical meaning. Hereafter, every selfadjoint operator will be called a (real-valued) observable.
 
 Let $\{X(t), t\in T\}$ be a commuting family of observables where $T$ is an index set. Such a family can be simultaneously diagonalized and hence it is possible to speak of observables of the form $\sum_{j=1}^{n}\, x_j\, X(t_j)$ for any finite set 
 $\{ t_1,t_2,\ldots t_n\}\subset T$ and real scalars $x_j, 1\leq j\leq n$. Define 
 \begin{equation}
 \label{1.4}
 \phi^u_{t_1,t_2,\ldots t_n}(\mathbf{x})=\langle u\vert e^{i\sum_{j=1}^n\, x_j\, X(t_j)}\vert u\rangle
 \end{equation}
  where $\mathbf{x}\in \mathbb{R}^n$ and $u$ is a fixed unit vector in $\mathcal{H}$. Then the left hand side of (\ref{1.4}) is the Fourier transform of a probability distribution $\mu^u_{t_1,t_2,\ldots, t_n}$ in $\mathbb{R}^n$ and the family $\{\mu^u_{t_1,t_2,\ldots, t_n}\vert \{t_1,t_2,\ldots, t_n\}\subset T, n=1,2,\ldots \}$ of all finite dimensional distributions is consistent. Thanks to Kolmogorov's theorem, one obtains a stochastic process. Such a view opens a quantum probabilistic route to constructing classical stochastic processes. Once again, it is interesting to note that the correspondence 
  \begin{equation*}
  (x_1,x_2,\ldots , x_n)\longrightarrow e^{i\sum_{j=1}^n\, x_j\, X(t_j)}
  \end{equation*} 
is a unitary representation of $\mathbb{R}^n$.   

In the discussion of the last paragraph, if $T$ is an interval in $\mathbb{R}$ and the observable-valued map $t\rightarrow X(t)$ is reasonably {\em smooth}, one can even look forward to a quantum stochastic operator differential description of $dX(t)$ in terms of a chosen family of operator processes. More of this in Section~4.

Suppose $G$ is a locally compact second countable group and $g\rightarrow U_g$ is a strongly continuous unitary representation of G in a Hilbert space $\mathcal{H}$. Let $A\subset G$ be any closed abelian subgroup and let $\hat{A}$ be the dual character group of $A$. Then there exists a spectral measure $P^{\hat{A}}$ such that 
\begin{equation*}
U_x=\int_{\hat{A}}\, P^{\hat{A}}(dy)\, \langle x, y\rangle, \ \ \ x\in A
\end{equation*}  
where  $\langle x, y\rangle$ is the value of the character $y$ at $x$. If $u$ is a fixed unit vector in $\mathcal{H}$, then the definition 
\begin{equation*}
\mu(F)=\langle u\vert P^{\hat{A}}(F)\vert u\rangle, \ \ \ F\subset \hat{A},\ \ {\rm a\ Borel\ subset}
\end{equation*}
yields a probability measure in $\hat{A}$ with the interpretation that $\mu$ is the probability distribution of an $\hat{A}$-valued observable $P^{\hat{A}}$. In other words, $\mu(F)$ is the probability of the event that the $\hat{A}$ valued observable takes a value in $F$. Thus varying $A$ and $u$ one obtains a rich class of observables and distributions. 

When the abelian subgroup $A$ happens to be a one parameter subgroup $t\rightarrow x_t$, $t\in \mathbb{R}$, then $\widetilde{U}_t=U_{x_t},\ t\in \mathbb{R}$ is a one paramter group of unitary operators in $\mathcal{H}$ and by Stone's theorem $\widetilde{U}_t=e^{itX}$ for all $t$, where $X$ is an observable. Many physically meaningful observables like energy, linear momenta, angular momenta, spin angular momenta etc., arise from the analysis of such observables. There are important commutation relations among them. They exercise a deep influence on quantum statistics. For an initiation into this world we recommend G. W. Mackey~\cite{Ma} and V.S. Varadarajan~\cite{Va}. 

The galilean group, inhomogeneous Lorentz group, SU(2), and U(3) are some examples which commonly occur in the description above. In our present exposition we shall deal with the euclidean group $\mathbb{E}(\mathcal{H})$ of a Hilbert space $\mathcal{H}$. It is the semidirect product of the additive group $\mathcal{H}$ and the unitary group $\mathcal{U}(\mathcal{H})$ of all unitary operators on $\mathcal{H}$, acting on $\mathcal{H}$. When $\mathcal{H}$ is infinite dimensional, 
 $\mathbb{E}(\mathcal{H})$   is non locally compact. However, $\mathbb{E}(\mathcal{H})$ admits a canonical, projective, unitary, irreducible and factorizable representation in the boson Fock space $\Gamma(\mathcal{H})$ over $\mathcal{H}$. We call it the {\em Weyl representation.} Its infinitesimal (Lie algebraic) version presents three fundamental operator fields called {\em creation, conservation} and {\em annihilation.} These are introduced in Section~3. 
 
 With a time parameter thrown in, one obtains the fundamental operator processes called creation, conservation and annihilation again. The notion  of a quantum stochastic integral with respect to a fundamental process becomes possible. This leads to a stochastic calculus with a quantum It{\^o}'s formula for products of differentials of fundamental processes and time. This is discussed briefly in Section 4. 
 
 As an application of the calculus in Section~4, we describe in the last section, the realization of all the classical L{\'e}vy processes with independent increments as linear combinations of the three  fundamental processes as well as an It{\^o}'s product formula for their differentials. 
  
\section{Notational preliminaries}

All the Hilbert spaces in this paper will be assumed to be complex and separable. The scalar product $\langle u\vert v\rangle$ between any two elements $u$ and $v$ in ${\mathcal H}$ will be assumed to be linear in the variable $v$ and antilinear in $u$. The symbol $\langle u \vert $ is considered as a linear functional on ${\mathcal H}$ and called a {\em bra} vector, whereas $\vert v\rangle$ is an element of ${\mathcal H}$, called a {\em ket} vector, so that $\langle u\vert$, evaluated at $\vert v\rangle$, denoted by the bracket $\langle u\vert v\rangle$ is the scalar product. In particular, $\langle v\vert u\rangle=\overline{\langle u\vert v\rangle}=\langle u\vert v\rangle^*$, the complex conjugate of $\langle u\vert v\rangle$.   The product $\vert v\rangle \langle u\vert$ is defined to be the operator satisfying 
\begin{equation*}
\vert v\rangle \langle u\vert\, \vert w\rangle = \langle u\vert w\rangle \, \vert v\rangle,\ \ \ \ \ \ \ \ \forall\  w\in \mathcal{H}. 
\end{equation*}
If $u\neq0,\ v\neq 0,$ the product operator $\vert v\rangle \langle u\vert$ is a rank one operator with the one dimensional range $\mathbb{C}\, v$. If $u=0$ or $v=0$, then  $\vert v\rangle \langle u\vert=0.$ If $u_j, \ v_j\in \mathcal{H},\, 1\leq j\leq n$, then 
\begin{equation*}
\prod_{j=1}^{n}\left(\vert v_j\rangle \langle u_j\vert\right)= 
\left(\prod_{j=1}^{n-1}\, \langle u_j\vert v_{j+1}\rangle \right) \, \vert v_1\rangle\langle u_n\vert. 
\end{equation*} 
We denote by $\mathcal{B}(\mathcal{H})$ the $*$ algebra of all bounded operators on $\mathcal{H}$ and the adjoint of an operator $A$ by $A^\dag$ instead of $A^*$. Any scalar multiple $c\, I$ of the identity operator $I$ will be denoted by the symbol $c$ itself. If $A$ is an operator in $\mathcal{H}$ and $c$ is a scalar, $A\pm cI=A\pm c.$ The multiplicative group of all unitary operators on $\mathcal{H}$ will be denoted by $\mathcal{U}(\mathcal{H})$. Then $\mathcal{U}(\mathcal{H})$ acts on the additive group $\mathcal{H}$ and determines the semidirect product 
\begin{equation}
\label{2.1}
\mathbb{E}(\mathcal{H})=\mathcal{H}\,\textcircled{s}\, \mathcal{U}(\mathcal{H}),
\end{equation}   
which, as a set, is the cartesian product $\mathcal{H}\times \mathcal{U}(\mathcal{H})$ but a group with the operation  
\begin{equation}
\label{2.2} 
(u, U)\, (v,V)=(u+Uv, U\,V)\ \ \forall \ \ u,v\in \mathcal{H},\ U,\, V \in \mathcal{U}(\mathcal{H}). 
\end{equation}
The norm topology of $\mathcal{H}$ and the strong topology in $\mathcal{U}(\mathcal{H})$ determine the product topology of $\mathbb{E}(\mathcal{H})$ under which the group operation (\ref{2.2}) is continuous. The topological group $\mathbb{E}(\mathcal{H})$ thus obtained, is called the {\em euclidean group} of $\mathcal{H}$. It is locally compact if $\mathcal{H}$ is finite dimensional, but not locally compact if ${\rm dim}\mathcal{H}=\infty$. This group will play an important role in our exposition. 

For any $A\in \mathcal{B}(\mathcal{H}), \ u, v\in \mathcal{H}$, we write 
	\begin{equation}
	\label{2.3} 
	\langle u\vert A\vert v\rangle=\langle u\vert A\, v\rangle=\langle A^\dag\, u\vert v\rangle. 
	\end{equation}
We have to deal with unbounded operators which are defined only on a linear manifold in $\mathcal{H}$. The linear manifold in $\mathcal{H}$ on which an operator $A$ is defined is called its domain, denoted by $\mathcal{D}_A$ or simply $\mathcal{D}$. Two such operators $A$, $B$ are said to be equal if $\mathcal{D}_A=\mathcal{D}_B$ and both $A$, $B$ agree on the domain. When such an operator $A$ admits an adjoint, the adjoint operator is denoted by $A^\dag$. For $u\in \mathcal{D}_{A^\dag}$ and 
$v\in \mathcal{D}_{A}$, relation (\ref{2.3}) holds. When $A=A^\dag$, we say that $A$ is selfadjoint and is called an {\em observable}. Every observable has a spectral representation as mentioned in (\ref{1.1}). 

We shall come across pairs of operators $A$, $B$, each with its own domain, satisfying the relation, $\langle u\vert A\, v\rangle=\langle B\, u\vert v\rangle$ for all $u, \, v$ in a dense linear manifold $\mathcal{D}\subset\mathcal{H}$. Thus we say that $A$ and $B$ are adjoint to each other in the domain $\mathcal{D}$ and abuse our notation to denote $B$ by $A^\dag$. 

If $\mathcal{H}_1,\ \mathcal{H}_2$ are two Hilbert spaces, we denote their Hilbert space tensor product by $\mathcal{H}_1\otimes\mathcal{H}_2$. If $\{u_i\},\ \{v_j\}$ are complete orthonormal bases in $\mathcal{H}_1,\,\mathcal{H}_2$ respectively, then the set $\{u_i\otimes v_j\}$ of all product vectors is a complete orthonormal basis in $\mathcal{H}_1\otimes\mathcal{H}_2$. A set $S$ in a Hilbert space 
$\mathcal{H}$ is said to be {\em total} if the smallest closed subspace containing $S$ is the whole space $\mathcal{H}$. If $S_i\subset\mathcal{H}_i$ is total for $\mathcal{H}_i$, then the set of product vectors $\{u\otimes v\vert u\in S_1, v\in S_2\}$ is total in $\mathcal{H}_1\otimes\mathcal{H}_2$. For any finite set $\{\mathcal{H}_i,\ 1\leq i\leq n\}$, one can define their tensor product $\mathcal{H}_1\otimes\mathcal{H}_2\otimes \ldots\otimes\mathcal{H}_n$. If  $\mathcal{H}_i=\mathcal{H}$ for all $i$, then we denote this product by $\mathcal{H}^{\otimes n}$. For any product vector 
$u_1\otimes u_2\otimes \ldots \otimes u_n$ and any permutation $\sigma$ of $\{1,2,\ldots, n\}$, we write 
$\sigma\, u_1\otimes u_2\otimes \ldots \otimes u_n=u_{\sigma(1)}\otimes u_{\sigma(2)}\otimes \ldots \otimes u_{\sigma(n)}$. Then there exists a unitary operator $U_\sigma$ on  $\mathcal{H}^{\otimes n}$, satisfying 
$U_\sigma\, u_1\otimes u_2\otimes \ldots \otimes u_n=u_{\sigma(1)}\otimes u_{\sigma(2)}\otimes \ldots \otimes u_{\sigma(n)}$ for all product vectors. Then the subspace 
\begin{equation*}
\{u\vert u\in \mathcal{H}^{\otimes n}, U_\sigma\, u=u\, \ \forall \sigma\in S_n \},
\end{equation*}
where $S_n$ is the permutation group of $\{1,2,\ldots , n\}$, called the $n$-fold symmetric tensor product of $\mathcal{H}$ and denoted by $\mathcal{H}^{\textcircled{s}^n}$. The Hilbert space $\Gamma(\mathcal{H})$ defined by the countable direct sum 
\begin{equation}
\label{2.4}
\Gamma(\mathcal{H})=\mathbb{C}\oplus \mathcal{H}\oplus \mathcal{H}^{\textcircled{s}^2}\oplus \ldots \oplus \mathcal{H}^{\textcircled{s}^n} \oplus \ldots 
\end{equation} 
is called the {\em boson Fock space} over $\mathcal{H}$. Here we shall view each $\mathcal{H}^{\textcircled{s}^n}$ as a subspace of $\Gamma(\mathcal{H})$ and call it the {\em $n$-particle subspace}. The one dimensional subspace $\mathbb{C}=\mathcal{H}^{\textcircled{s}^0}$ of all complex scalars is called the {\em vacuum} subspace. Any element of $\mathcal{H}^{\textcircled{s}^n}$ is called an $n$-{\em particle vector}. Any element of the form $u_0\oplus u_1\oplus \ldots\oplus u_n$ where each $u_j$ is a $j$-particle vector and $n$ is arbitrary, is called a {\em finite particle vector}. The dense linear manifold $\mathcal{F}(\mathcal{H})$ of all finite particle vectors is called the {\em finite particle domain}. 

For any $u\in \mathcal{H}$, we write 
\begin{equation} \label{2.5}
e (u) = 1 \oplus u \oplus \frac{u^{\otimes^{2}}}{\sqrt{2!}}  \oplus \cdots 
\oplus \frac{u^{\otimes^{r}}}{\sqrt{r!}}  \oplus \cdots , \end{equation}
and call it the exponential vector associated with $u$. The vector $e(0)$ is called the {\em vaccum vector}. Exponential vectors satisfy the following properties: 
\begin{enumerate}[label=(\roman*)]
	\item For all $u, v\in \mathcal{H}$, we have
	\begin{equation}
	\label{2.6}
	\langle e(u)\vert e(v)\rangle=e^{\langle u\vert v\rangle}. 
	\end{equation}
	\item For any finite set $F=\{u_1,u_2,\ldots, u_n\}\subset \mathcal{H}$, the set 
	$\{e(u_1),e(u_2),\ldots , e(u_n)\}\subset\Gamma(\mathcal{H})$ is linearly independent. 
	\item The map $u\rightarrow e(u)$ from $\mathcal{H}$ into  $\Gamma(\mathcal{H})$ is continuous. 
	\item $\{e(u) \vert u\in \mathcal{H}\}$ is total in $\Gamma(\mathcal{H})$. 
	\item If T is a contraction operator on $\mathcal{H}$, so is $T^{\otimes n}$ on the $n$-particle space $\mathcal{H}^{\textcircled{s}^n}$ for every $n$. Thus the operator $\Gamma(T)$, called the {\em second quantization} of $T$, defined by 
	 \begin{equation*}
	 \Gamma(T)=I\oplus T\oplus T^{\otimes 2}\oplus \ldots \oplus T^{\otimes n}\oplus \ldots
	 \end{equation*}
	 on $\Gamma(\mathcal{H})$ is again a contraction operator satisfying 
	 \begin{equation*}
	 \Gamma(T)\,e(u)=e(Tu)\, \ \ \forall u\in \mathcal{H}.  
	 \end{equation*}
	 The identity-preserving correspondence $T\rightarrow \Gamma(T)$, called the {\em second quantization map}, satisfies the following: 
	 \begin{eqnarray*}
	 	\Gamma(T^\dag)&=&\Gamma(T)^\dag\\ 
	 	\Gamma(T_1\,T_2)&=&\Gamma(T_1)\, \Gamma(T_2).
	 \end{eqnarray*}
	 Thus, $\Gamma(T)$ is a self adjoint, positive, projection, isometric, coisometric or unitary operator according as $T$ is. In particular, the map 
	 $U\rightarrow \Gamma(U)$ on the unitary group $\mathcal{U}(\mathcal{H})$ is a strongly continuous unitary representation which leaves each $n$-particle subspace invariant. 
	 \item If $H$ is an observable in $\mathcal{H}$, then $\{e^{-itH}, t\in \mathbb{R}\}$ is a one parameter unitary group and $\{\Gamma(e^{-itH}), t\in \mathbb{R}\}$ is the corresponding second quantized one parameter unitary group in $\Gamma(\mathcal{H})$. Its Stone generator $\lambda(H)$ satisfying 
	 \begin{equation*}
	 e^{-it\,\lambda(H)}=\Gamma(e^{-itH})
	 \end{equation*}
	 is an observable called the {\em differential second quantization} of $H$. 
	 \item If $\mathcal{H}_i, \ 1\leq i\leq n$ are Hilbert spaces, $u_i\in\mathcal{H}_i,\ 1\leq i\leq n$, then the correspondence 
	 \begin{equation*}
	 e(u_1)\otimes e(u_2)\otimes\ldots \otimes e(u_n)\rightarrow e(u_1\oplus u_2\oplus\ldots \oplus u_n)   
	 \end{equation*}
	 between products of exponential vectors in $\Gamma(\mathcal{H}_i), \ 1\leq i\leq n$ and exponential vectors in  
	 $\Gamma(\oplus_1^n\, \mathcal{H}_i)$ is scalar product preserving and therefore extends to a natural Hilbert space isomorphism between $\otimes_1^n\,\Gamma(\mathcal{H}_i)$ and $\Gamma(\oplus_1^n\, \mathcal{H}_i)$.   We shall often identify these two Hilbert spaces and write 
	 \begin{eqnarray*}
	 e(u_1)\otimes e(u_2)\otimes\ldots \otimes e(u_n)&=& e(u_1)\,e(u_2)\, \ldots e(u_n) \\ 
	 &=& e(u_1\oplus u_2\oplus\ldots \oplus u_n).    
	 \end{eqnarray*} 
 If $P(\cdot)$ is a continuous spectral measure on $\mathbb{R}_+=[0,\infty)$ with respect to a Hilbert space $\mathcal{H}$ and $\mathcal{H}([a,b])$ is the range of $P([a,b])$ then we have 
 \begin{equation*}
 \Gamma(\mathcal{H})=\Gamma(\mathcal{H}([0,t_1]))\otimes \Gamma(\mathcal{H}([t_1,t_2])) \otimes \ldots \otimes \Gamma(\mathcal{H}([t_{n-1},t_n]))\otimes \Gamma(\mathcal{H}([t_{n},\infty))
\end{equation*}
	for any partition $0<t_1<t_2< \ldots t_n<\infty$ of $\mathbb{R}_+.$ Thus $\Gamma(\mathcal{H})$ exhibits {\em `continuous tensor product property'}. For more details and applications see Ref.~\cite{KRP}. 
\end{enumerate}  

 \noindent{\bf Example 2.1}\ Consider the Hilbert spaces $L^2(\mathbb{R}_+), \Gamma(L^2(\mathbb{R}_+))$ and $L^2(\mu)$, where $\mu$ is the Wiener probability measure $\mu$ of the standard brownian motion $\{B(t), t\geq 0\}$. For any $u\in L^2(\mathbb{R}_+)$ consider the correspondence 
 \begin{equation*}
 e(u)\rightarrow \widetilde{e}(u)(B)= {\rm exp}\,\left( \int_0^\infty u\, dB- 
 \frac{1}{2}\,\int_0^\infty\, u(t)^2\, dt\right),\ \ u\in L^2(\mathbb{R}_+), 
 \end{equation*} 
 Then 
 \begin{equation*}
 \langle e(u)\vert\, e(v)\rangle={\rm exp}\langle u\vert v\rangle=\mathbb{E}_\mu\, \overline{\widetilde{e}(u)}\, \widetilde{e}(v),\ \ \ \forall\ \ u, v\in L^2(\mathbb{R}_+). 
 \end{equation*}
 Thus the correspondence described above extends uniquely to a unitary isomorphism between the boson Fock space over $\Gamma(L^2(\mathbb{R}_+))$ and the Hilbert space $L^2(\mu)$. Under this correspondence, $e(\mathbbm{1}_{[0,t]}\,u)$ goes to 
 ${\rm exp}\,\left( \int_0^t u\, dB- 
 \frac{1}{2}\,\int_0^t\, u(s)^2\, ds\right)$ for all $t\geq 0$. If $P([0,t])$ is the projection of multiplication by $\mathbbm{1}_{[0,t]}$ in $L^2(\mathbb{R}_+)$, then $\Gamma(P([0,t]))$ goes to the conditional expectation, given the brownian path upto time $t$. 
 
 
 We denote by $\mathcal{E}(\mathcal{H})$ the linear manifold generated by the set 
 $\{e(u), u\in \mathcal{H}\}$ of all exponential vectors and call it the {\em exponential domain} in $\Gamma(\mathcal{H})$. By property (iv) above, $\mathcal{E}(\mathcal{H})$ is dense in $\Gamma(\mathcal{H})$. If $T$ is any map from the set of all exponential vectors into $\Gamma(\mathcal{H})$, it follows from property (ii) that $T$ extends uniquely linearly to an operator in $\Gamma(\mathcal{H})$ with the dense domain $\mathcal{E}(\mathcal{H})$. 
 
 \section{The Weyl representation and the fundamental operator fields in $\Gamma(\mathcal{H})$}
 
 We begin with the construction of the Weyl representation of the euclidean group $\mathbb{E}(\mathcal{H})$ of the Hilbert space $\mathcal{H}$, from which we derive the basic observables needed in this article. The additive subgroup $\mathcal{H}$ of  $\mathbb{E}(\mathcal{H})$ has a natural translation action on the set of all exponential vectors by $u:\, e(v)\rightarrow e(v+u)$, but it is not scalar product preserving. By introducing a scalar multiplication factor, we define
 \begin{equation}
 \label{3.1}
 W(u) e(v) = e^{-\frac{1}{2} ||u||^{2} - \langle u|v \rangle} 
 e (u+v) \quad \forall\, \ v\in\mathcal{H}.
 \end{equation} 
 It follows from (\ref{2.6}) that 
 \begin{equation*}
 \langle W(u)\,e(v_1)\vert W(u)\, e(v_2)\rangle= \langle e(v_1)\vert e(v_2)\rangle
 \end{equation*}
 for all $v_1,\ v_2$ in $\mathcal{H}$. The totality of the set of all exponential vectors in $\Gamma(\mathcal{H})$ implies the existence of a unique unitary operator $W(u)$ in $\Gamma(\mathcal{H})$ satisfying (\ref{3.1}) for every $u$ in $\mathcal{H}$. Successive applications on the exponential vectors shows that 
 \begin{equation}
 \label{3.2}
 W(u_1) W(u_2) = e^{-i\, {\rm Im}\, \left( \langle u_1|u_2 \rangle\right)} \, W(u_1+u_2). 
 \end{equation}
  This also implies 
 \begin{equation}
 \label{3.3}
 W(u_1) W(u_2) = e^{-2\,i\, {\rm Im}\, \left( \langle u_1|u_2 \rangle\right)} \, W(u_2)\, W(u_1). 
 \end{equation} 
 These are known as {\em Weyl commutation relations}. 
 
 In Section 2, we have already noted the action of the unitary group $\mathcal{U}(\mathcal{H})$ on the exponential vectors: $\Gamma(U)\, e(v)=e(U\,v), 
 U\in  \mathcal{U}(\mathcal{H})$ yielding the second quantization representation. Combining this with the operator $W(u)$, satisfying (\ref{3.1}), we define 
 \begin{equation}
 \label{3.4}
 W(u,U)=W(u)\, \Gamma(U), \ \ (u,U)\in \mathbb{E}(\mathcal{H}).
 \end{equation}
 Once again, by verification on exponential vectors, we get the relations 
 \begin{eqnarray}
 \label{3.5}
 \Gamma(U)\, W(u)\, \Gamma(U)^{-1}&=&W(U\, u), \nonumber \\ 
  W(u_1,U_1)\,  W(u_2,U_2)&=&e^{-i\, {\rm Im}\, \left( \langle u_1|U_1\,u_2\rangle\right)}\, W(u_1+U_1\,u_2, U_1U_2) 
 \end{eqnarray}
 for all $u,u_1,u_2$ in $\mathcal{H}$ and $U, \, U_1,\ U_2$ in $\mathcal{U}(\mathcal{H})$. Furthermore, the map $(u,\, U)\rightarrow W(u,U)$ from 
 $\mathbb{E}(\mathcal{H})$ into $\mathcal{U}(\Gamma(\mathcal{H}))$ is strongly continuous. In other words, we have obtained a projective unitary representation of the euclidean group $\mathbb{E}(\mathcal{H})$ in $\Gamma(\mathcal{H})$. We call this representation, the {\em  Weyl representation} of $\mathbb{E}(\mathcal{H})$.
 	
 	If $\mathcal{H}=\mathcal{H}_1\oplus\mathcal{H}_2$, we have 
 	$\Gamma(\mathcal{H})=\Gamma(\mathcal{H}_1)\otimes \Gamma(\mathcal{H}_2)$ by the identification described in property (vii) of Section~2 and for any $u_i\in\mathcal{H}_i,\, U_i\in \mathcal{U}(\mathcal{H}_i),\ i=1,2$, \ 
 	\begin{equation*}
 	W(u_1\oplus u_2,\, U_1\oplus U_2)\, e(v_1\oplus v_2)= \{ W(u_1,U_1)\,e(v_1)\}\otimes \{ W(u_2,U_2)\,e(v_2)\} 
 	\end{equation*} 
 	for all $v_i\in\mathcal{H}_i,\ i=1,2.$ In other words 
 	\begin{equation*}
 	W(u_1\oplus u_2,\, U_1\oplus U_2)=  W(u_1,U_1)\,\otimes W(u_2,U_2). 
 	\end{equation*} 
 	We say that the Weyl representation has the {\em factorizability property} in the context of tensor products. 
 	
 	The projective representation $u\rightarrow W(u)$ satisfying (\ref{3.1}) and 
 	(\ref{3.2}) is also irreducible in $\Gamma(\mathcal{H})$, i.e., there is no proper subspace in $\Gamma(\mathcal{H})$, invariant under all $W(u),\ u\in\mathcal{H}$. Thus the Weyl representation is also irreducible. For a proof of this theorem we refer to Chapter~II, Section~20 in Ref.~\cite{KRP}.  
 
 From (\ref{3.2}) it follows that the map $t\rightarrow W(tu)$, $t\in \mathbb{R}$ is a one parameter group of unitary operators, which has a selfadjoint Stone generator $p(u)$ satisfying the relation 
 \begin{equation}
 \label{3.6} 
 W(t\,u)=e^{-it\, p(u)},\  u\in\mathcal{H}
 \end{equation} 
 yielding the family $\{p(u),\ u\in\mathcal{H}\}$ of observables in $\Gamma(\mathcal{H})$, corresponding to the translation subgroup $\mathcal{H}$ in 
 $\mathbb{E}(\mathcal{H})$. 
 
 As we have already observed in property (vi) of Section~2, to each observable $H$ in $\mathcal{H}$, its differential second quantization $\lambda(H)$ is an observable in $\Gamma(\mathcal{H})$ satisfying 
 \begin{equation}
 \label{3.7}
 \Gamma(e^{-itH})=e^{-it\, \lambda(H)},\ H {\rm\ an \ observable\ in\ }\mathcal{H}. 
 \end{equation}
 The family of observables $\{\lambda(H), \, H\ $ an observable in $\mathcal{H}\}$ accounts for the unitary subgroup $\mathcal{U}(\mathcal{H})$ in $\mathbb{E}(\mathcal{H})$. 
 
 The observables $p(u)$, $u\in \mathcal{H}$ and $\lambda(H),\ H$ an observable in $\mathcal{H}$, together yield an infinitesimal or Lie algebraic version of the Weyl representation. Going back to equation (\ref{3.1}), replacing $u$ by $t\, u,\ t\in \mathbb{R}$ and differentiating with respect to $t$ at 0, one gets from (\ref{3.6}) 
 \begin{equation*}
 p(u)\, e(v)=-i\,\langle u\vert v\rangle\,e(v)\, +\, i\, \sum_{n=1}^{\infty}\frac{1}{\sqrt{n !}} \sum_{r=0}^{n-1}\, v^{\otimes^r}\otimes u\otimes v^{\otimes^{n-r-1}}. 
 \end{equation*}
 Define the operators 
 \begin{eqnarray}
 \label{3.8}
 a(u)&=&\frac{p(-iu)+i\,p(u)}{2}, \\
 \label{3.9}
 a^\dag(u)&=&\frac{p(-iu)-i\,p(u)}{2}, \  \ u\in \mathcal{H}
 \end{eqnarray}
 and observe that 
 \begin{eqnarray}
 \label{3.10}
 a(u)\, e(v)&=&\langle u\vert v\rangle\, e(v), \\
 \label{3.11}
 a^\dag(u)\, e(v)&=&\sum_{n=1}^{\infty}\frac{1}{\sqrt{n !}} \sum_{r=0}^{n-1}\, v^{\otimes^r}\otimes u\otimes v^{\otimes^{n-r-1}}.
 \end{eqnarray}
 Changing $v$ to $s\, v,\ s\in\mathbb{R}$, and identifying coefficients of $s^n$ on both sides of these equations, we get 
 \begin{eqnarray}
 \label{3.12}
 a(u)\, v^{\otimes^n}&=&\sqrt{n}\, \langle u\vert v\rangle\, v^{\otimes^{n-1}}, \\
 \label{3.13}
 a^\dag(u)\, v^{\otimes^n}&=&\frac{1}{\sqrt{n+1}}\sum_{r=0}^{n} \, v^{\otimes^r}\otimes u\otimes v^{\otimes^{n-r}}.
 \end{eqnarray}
 for all $v\in\mathcal{H}$. It may also be noted that 
 \begin{equation}
 \label{3.14}
 a^\dag(u)\, e(v)=\left.\frac{d}{ds}\, e(v+s\, u)\right\vert_{s=0}.
 \end{equation}  
 Equations (\ref{3.8})--(\ref{3.14}) lead to the following facts: 
 
 Both $\mathcal{E}(\mathcal{H})$ and $\mathcal{F}(\mathcal{H})$, namely the exponential and finite particle domains are contained in the domains of $a(u)$ and $a^\dag(u)$ for every $u$ in $\mathcal{H}$. The operators $a(u)$ and $a^\dag(u)$ are adjoint to each other in these domains. Exponential vectors are eigenvectors for every $a(u)$. The operator $a(u)$ maps the $n$-particle subspace into the $(n-1)$-particle subspace and annihilates the vacuum vector for every $u\neq 0$. The operator $a^\dag(u)$ maps the $n$-particle subspace into  the $(n+1)$-particle subspace and the vacuum vector $e(0)$ to the 1-particle vector $u$ for $u\neq 0$. On $\mathcal{E}(\mathcal{H})$ and $\mathcal{F}(\mathcal{H})$, the following commutation relations hold: 
 \begin{eqnarray}
 \label{3.15}
 [a(u)\, , a(v)]&=&0,\  \ [a^\dag(u)\, , a^\dag(v)]=0 \\
 \label{3.16}
 [a(u)\, , a^\dag(v)]&=& \langle u\vert v\rangle,\ \ u,\ v \in\mathcal{H}. 
 \end{eqnarray}
 The map $u\rightarrow a(u)$ is antilinear, whereas $u\rightarrow a^\dag(u)$ is linear. The operator $a(u)$ is called the {\em annihilation operator} associated with $u$ and $a^\dag(u)$, the {\em creation operator}, associated with the same $u$. 
 
 The observables $p(u)$ defined by (\ref{3.6}) satisfy 
 \begin{equation}
 \label{3.18}
 p(u)=\,i\, \left(a^\dag(u)-a(u)\right) 
 \end{equation}
 on the domains $\mathcal{E}(\mathcal{H})$ and $\mathcal{F}(\mathcal{H})$. Both these domains are cores for $p(u)$ for every $u$. 
 
 Choose and fix a complete orthonormal basis $\{e_r,\ r=1,2,\ldots\}$ and define 
 \begin{equation*}
 p_r=c\, p(e_r),\ q_r=-\frac{1}{2c}\, p(i\,e_r).
 \end{equation*} 
 where $c$ is a positive scalar. Then (\ref{3.15}) and (\ref{3.16}) imply that on each of the domains $\mathcal{E}(\mathcal{H})$, $\mathcal{F}(\mathcal{H})$, the observables $q_1,q_2,\ldots $ and $p_1,p_2,\ldots$ satisfy the commutation relations 
 \begin{eqnarray}
 \label{3.18}
 [p_r\, , p_s]&=&0,\  \ [q_r\, , q_s]=0, \\
 \label{3.19}
 [q_r\, , p_s]&=& i\,\delta_{r\,s}, 
 \end{eqnarray}
 for all $r,\,s$ in $\{1,2,\ldots\}$. These are the position-momentum commutation relations of Heisenberg in quantum theory, which are realized as the Lie algebraic version of the Weyl representation of the group $\mathbb{E}(\mathcal{H})$, but using only the additive subgroup of $\mathcal{H}$. They are usually called the canonical commutation relations or CCR in the form (\ref{3.15}) and (\ref{3.16}) or 
 (\ref{3.18}) and (\ref{3.19}). 
 
 We now turn to the unitary subgroup $\mathcal{U}(\mathcal{H})$ in $\mathbb{E}(\mathcal{H})$. Recall from property (vi) of Section~2, that corresonding to any observable $H$ in $\mathcal{H}$, there is a one parameter unitary subgroup $\{U_t=e^{-it\,H}, t\in\mathbb{R}\}$, and $\{\Gamma(U_t), 
 t\in \mathbb{R}\}$ is a one parameter unitary group in $\Gamma(\mathcal{H})$ with its Stone generator $\lambda(H)$ satisfying 
 \begin{equation}
 \label{3.20}
 \Gamma(e^{-it\, H})=e^{-it\, \lambda(H)},\ t\in \mathbb{R}
 \end{equation}
 and 
 \begin{equation*}
 e^{-it\, \lambda(H)}\, e(v)=e(e^{-it\, H}\, v),\ v\in \mathcal{H}.
 \end{equation*}
 When $v$ is in the domain of $H$ we can differentiate at $t=0$ and obtain the relation 
 \begin{eqnarray}
 \label{3.21}
 \lambda(H)\, e(v)&=&\sum_{n=1}^{\infty} \, \frac{1}{\sqrt{n!}}\, 
 \sum_{r=0}^{n-1}\, v^{\otimes^r}\otimes H\, v\otimes v^{\otimes^{n-r-1}},  \\ 
 \label{3.22}
  \lambda(H)\, v^{\otimes^n}&=&\sum_{r=0}^{n-1} \, v^{\otimes^r}\otimes H\, v\otimes v^{\otimes^{n-r-1}}.
 \end{eqnarray}
 When $H$ is a bounded observable, the two relations above hold for every $v$ in $\mathcal{H}$. Thus both domains $\mathcal{E}(\mathcal{H})$ and $\mathcal{F}(\mathcal{H})$ are contained in the domain of the observable $\lambda(H)$. Note that $\lambda(H)$ leaves the $n$-particle subspace invariant. It also follows that for any two bounded observables $H_1$, $H_2$ in $\mathcal{H}$, 
 \begin{eqnarray}
 \label{3.23}
 \lambda(\alpha\, H_1+\beta\, H_2)&=&\alpha\, \lambda(H_1)+\beta\, \lambda(H_2),\ \ \alpha, \, \beta\in\mathbb{R}\\ 
 \lambda\left([i\,[H_1,H_2]\right)&=&i\, \left[\lambda(H_1),\, \lambda(H_2)\right]\nonumber
 \end{eqnarray}
 on the two domains $\mathcal{E}(\mathcal{H})$ and $\mathcal{F}(\mathcal{H})$.  
 
 For any bounded operator $K$ in $\mathcal{H}$, define 
 \begin{equation}
 \label{3.24}
 \lambda(K)=\lambda\left(\frac{K+K^\dag}{2}\right)+i\, \lambda\left(\frac{K-K^\dag}{2i}\right), \ \ K\in\mathcal{B}(\mathcal{H}).
 \end{equation}
 We call $\lambda(K)$ the {\em conservation} operator associated with $K$. The map 
 $K\rightarrow \lambda(K)$ with domain $\mathcal{E}(\mathcal{H})$ or $\mathcal{F}(\mathcal{H})$ is linear. In order to express the commutation relations among the three fundamental field operators of creation, conservation and annihilation, it is convenient to introduce the following notation inspired by that of Dirac: 
 $$a(\langle u\vert)=a(u), \ a^\dag(\vert u\rangle)=a^\dag (u),$$
 and $\lambda(K)$ is as it is, but keeping in mind operators of the form $K=\vert u\rangle \langle v\vert. $ Then, on the domains $\mathcal{E}(\mathcal{H})$ and $\mathcal{F}(\mathcal{H})$, the following hold: 
 \begin{eqnarray*}
 \left[ a^\dag(\vert u\rangle ), a^\dag(\vert v\rangle)\right]&=&0,\ \ \left[ a(\langle u\vert ), a(\langle v\vert)\right]=0,\\ 
 \left[ a(\langle u\vert ), a^\dag(\vert v\rangle)\right]&=&\langle u\vert v\rangle,\ \ \left[a(\langle u\vert ), \lambda(K )\right]=a(\langle u\vert \, K), \\ 
  \left[ \lambda(K ),  a^\dag(\vert v\rangle)\right]&=& a^\dag(K\,\vert v\rangle), \  \left[ \lambda(K_1 ), \lambda(K_2)\right]=\lambda([K_1,K_2]).
 \end{eqnarray*}
where $u,v\in\mathcal{H}$, $K, K_1,K_2\in \mathcal{B}(\mathcal{H})$. 

We call the set of all the relations above CCR. For their complete derivation with all the details, we refer the reader to Chapter~II, Section~2.3 in Ref.~\cite{KRP}. 

We now conclude the section with two important statistical features of the observables arising from the Weyl representation. From (\ref{3.1}) we have 
\begin{equation}
\label{3.25} 
\langle e(0)\vert W\left(\sum_{j=1}^{n} t_j\, u_j\,\right)\, \vert e(0)\rangle= 
{\rm exp}\, \left(-\frac{1}{2}\sum_{i,j}\,t_i\,t_j\, \langle u_i\vert u_j\rangle\, \right)
\end{equation} 
for $u_j\in \mathcal{H}, \ t_j\in \mathbb{R},\ 1\leq j\leq n$.  If $\mathcal{H}_\mathbb{R}$ is a real subspace of $\mathcal{H}$ such that 
$\mathcal{H}=\mathcal{H}_\mathbb{R}\oplus i\, \mathcal{H}_\mathbb{R}$, then $\langle u\vert v\rangle$ is real for $u,v$ in $\mathcal{H}_\mathbb{R}$ and by (\ref{3.3}), 
$\{W(u), u\in\mathcal{H}_\mathbb{R}\}$ is a commutative family of operators. By 
(\ref{3.6}), $\{p(u),u\in \mathcal{H}_\mathbb{R} \}$ is a commutative family of observables, which satisfies 
\begin{eqnarray*}
	\langle e(0)\vert W\left(\sum_{j=1}^{n} t_j\, u_j\,\right)\, \vert e(0)\rangle&=& 
	\langle e(0)\vert {\rm exp}\left(-i\sum_{j=1}^{n}\, t_j\, p(u_j)\,\right)\, \vert e(0)\rangle\\ 
	&=& {\rm exp}\, \left(-\frac{1}{2}\sum_{i,j}\,t_i\,t_j\, \langle u_i\vert u_j\rangle\, \right) ,\ \ t_j\in \mathbb{R}, \ 1\leq j\leq n,
\end{eqnarray*}
which is the characteristic function of the $n$-dimensional normal or Gaussian distribution with zero means and covariance matrix $((\langle u_i\vert u_j
\rangle ))$. In other words, the commutative family $\{p(u), u\in \mathcal{H}_\mathbb{R}\}$ executes a zero mean classical Gaussian random field in the vacuum state $\vert e(0)\rangle$. For the same reasons, the family 
$\{p(i\,u), u\in \mathcal{H}_\mathbb{R}\}$ also has the same property but
\begin{eqnarray*}
\left[p(u),\ p(i\,v)\right]\ &=&\ 2\, i\, \langle u\vert v\rangle, \ \ u, v\in \mathcal{H}_{\mathbbm{R}}, \\
\Gamma(i)\, p(u)\, \Gamma(-i)&=&p(i\,u).  
\end{eqnarray*}
Let now $u\in \mathcal{H}$, $U\in \mathcal{U}(\mathcal{H})$. Then 
\begin{eqnarray*}
\langle e(0)\vert W(-u)\, \Gamma(U)\, W(u)\, \vert e(0)\rangle &=& 
e^{-\vert\vert u\vert\vert^2}\, \langle e(u)\vert \, \Gamma(U)\,\vert e( u)\rangle \\ 
&=& e^{-\vert\vert u\vert\vert^2}\, \langle e(u)\vert \, e(U\, u)\rangle  \nonumber \\ 
&=& {\rm exp}\left(\langle u\vert U-I\vert u\rangle\right). 
\end{eqnarray*}
Choosing $U=U_t=e^{-i\,t\,H}$ where $H$ is an observable with spectral measure $P^H$, 
we have, 
\begin{eqnarray}
\label{3.26}
 \langle e(0)\vert W(-u)\, e^{i\, t\, \lambda(H)}\, W(u)\, \vert e(0)\rangle &=& 
 {\rm exp}\left(\langle u\vert e^{i\, t\, H}-I\vert u\rangle\right) \nonumber \\ 
& =& {\rm exp}\left(\int_\mathbb{R}\, (e^{itx}-1)\, \langle u\vert P^H\,(dx)\, \vert u\rangle \right), \ t\in\mathbb{R}, 
 \end{eqnarray}
 which is the characteristic function of the infinitely divisible distribution with L{\'e}vy measure $\langle u\vert P^H\,(\cdot)\, \vert u\rangle$. More generally, for any finite set of commuting observables $H_j$, $1\leq j\leq n$, the observables 
 $W(-u)\,\lambda(H_j)\, W(u)$, $1\leq j\leq n$ is also a commuting set of observables with a joint distribution in $\mathbb{R}^n$, which is infinitely divisible with L{\'e}vy measure $\langle u\vert P^{H_1,H_2,\ldots , H_n}(\cdot )\, \vert u\rangle$, where $P^{H_1,H_2,\ldots , H_n}$ is the joint spectral measure in $\mathbb{R}^n$ of $H_1, H_2, \ldots , H_n.$
 
 In short, the whole set $\left\{ W(-u)\, \lambda(H)\, W(u), \ H {\rm \ an\ observable \ in \  } \mathcal{H} \right\}$, is a non-commutative L{\'e}vy field, where restriction to  any real linear space of commuting elements  is a classical L{\'e}vy field in the vaccum state. 

Since the fundamental operator fields of creation, conservation and annihilation are linear combinations of the observables $p(u)$, \ $\lambda(H)$, where $u\in \mathcal{H}$ and $H$ is an observable in $\mathcal{H}$, the statistical properties embodied in (\ref{3.25}) and (\ref{3.26}) suggest the possibility of stochastic integration, if an appropriate time parameter is introduced in the fundamental fields. 
\section{Quantum stochastic integration and It{\^o}'s formula} 

In order to make room for the time parameter in the fundamental operator fields of the last section, we assume that the Hilbert space has the form, 
\begin{equation}
\label{4.1}
\mathcal{H}=\mathcal{K}\otimes L^2(\mathbb{R}_+)
\end{equation}
where $\mathbb{R}_+$ is the interval $[0,\infty)$ with its Borel structure and Lebesgue measure. We view $\mathcal{H}$ as the space of $\mathcal{K}$-valued norm square integrable functions, with the scalar product 
\begin{equation}
\label{4.2}
\langle f\vert g\rangle = \int_0^\infty\, \langle f(t)\vert g(t)\rangle_{\mathcal{K}}\, dt
\end{equation}
and drop the suffix $\mathcal{K}$ in the integrand. Consider the canonical spectral measure $P(\cdot)$ on $\mathbb{R}_+$ defined by 
\begin{equation*}
\left(P(E)f\right)(t)=\mathbbm{1}_E(t)\, f(t), \ \ f\in \mathcal{H},  
\end{equation*}
with $\mathbbm{1}_E$ denoting the indicator function of a Borel subset $E$ of $\mathbb{R}_+$. 

By property (vii) of Section 2, the boson Fock space $\widetilde{\mathcal{H}}=\Gamma(\mathcal{H})$ has a continuous tensor product structure over $\mathbb{R}_+$. For $0<s<t<\infty$ we write $\mathcal{H}_{s]}=\mathcal{K}\otimes L^2([0,s])),$ $\widetilde{\mathcal{H}}_{s]}=\Gamma(\mathcal{H}_{s]}),$ 
$\mathcal{H}_{[t}=\mathcal{K}\otimes L^2([t,\infty))),$ 
$\widetilde{\mathcal{H}}_{[t}=\Gamma(\mathcal{H}_{[t}),$ 
$\mathcal{H}_{[s,t]}=\mathcal{K}\otimes L^2([s,t])),$ 
$\widetilde{\mathcal{H}}_{[s,t]}=\Gamma(\mathcal{H}_{[s,t]})$ and note that 
\begin{eqnarray*}
\widetilde{\mathcal{H}}&=&\widetilde{\mathcal{H}}_{s]}\otimes \widetilde{\mathcal{H}}_{[s,t]}\otimes \widetilde{\mathcal{H}}_{[t},\\  
\widetilde{\mathcal{H}}_{[s,u]}&=&\widetilde{\mathcal{H}}_{[s,t]}\otimes \widetilde{\mathcal{H}}_{[t,u]},\  {\rm \ for\ } s<t<u.   
\end{eqnarray*} 
Denote the identity operators
in $\widetilde{\mathcal{H}}_{s]}$, $\widetilde{\mathcal{H}}_{[s,t]}$, $\widetilde{\mathcal{H}}_{[t}$ respectively by 
$\mathbbm{1}_{s]}$, $\mathbbm{1}_{[s,t]}$, $\mathbbm{1}_{[t}$. 

To begin with, we introduce the notion of a bounded operator valued adapted process in 
$\widetilde{\mathcal{H}}$. A map $t\rightarrow X(t)$ from $\mathbb{R}_+$ into $\mathcal{B}(\widetilde{\mathcal{H}})$ is called an adapted process, if there exist operators $X_t$ in $\widetilde{\mathcal{H}}_{t]}$ such that $X(t)=X_t\otimes \mathbbm{1}_{[t}$ for all $t$ and for any $\psi\in \widetilde{\mathcal{H}}$, the map $t\rightarrow X(t)\, \psi$ is measurable.  

 We have to deal with unbounded operators here. For simplicity of exposition we assume that all the operators in any boson Fock space that appear hereafter in this paper have domains which include the exponential domain. For any $f$ in $\mathcal{H}$ we write 
 \begin{eqnarray*}
 	f_{s]}=P([0,s])\, f, \ \   f_{[s,t]}=P([s,t])\, f,\ \ f_{[t}&=&P([t,\infty))\, f
 \end{eqnarray*}
and note that 
 \begin{eqnarray*}
 e(f)&=&e(f_{s]})\, e(f_{[s,t]})\, e(f_{[t}),\\ 
 e(f_{[s,u]})&=&e(f_{[s,t]})\, e(f_{[t,u]}),\ \ {\rm \ for\ all\ \ } 0<s<t<u<\infty.   
\end{eqnarray*}
A map $t\rightarrow X(t)$ from $\mathbb{R}_+$ into the space of operators in $\widetilde{\mathcal{H}}$ is called an adapted process if there exists an operator $X_t$ in 
$\mathcal{H}_{t]}$ such that 
\begin{equation}
\label{4.3}
X(t)\, e(f)= \left( X_t\, e(f_{t]}\right)\otimes e(f_{[t}), \ \ f\in \mathcal{H}, t\in \mathbb{R}_+
\end{equation}  
and the map $t\rightarrow X(t)\, e(f)$ is measurable for every $f$ in $\mathcal{H}$. Such an adapted process is called {\em regular} if the map $t\rightarrow X(t)\, e(f)$ is continuous for every $f$ in $\mathcal{H}$. 

An adapted process $t\rightarrow X(t)$ is said to be {\em simple} if there exist $0<t_1<t_2<\ldots <t_n<\ldots ,\ \ t_n\rightarrow \infty$ as $n\rightarrow \infty$ and 
\begin{equation*}
X(t)=\sum_{j=0}^\infty X(t_j)\, \mathbbm{1}_{[t_j,t_{j+1})}(t)\ \ \forall \ t
\end{equation*} 
  where $t_0=0$ and the interval $[t_j,t_{j+1})$ is closed at $t_j$ and open at $t_{j+1}$. In other words 
  \begin{equation*}
  X(t)= X(t_j)\, \ {\rm if}\  t_j\leq t<t_{j+1}  \ \forall \ j.
  \end{equation*} 
  Denote by $\mathcal{H}_{\rm loc}$ the space of all Borel maps $\phi:\ \mathbb{R}_+\rightarrow\, \mathcal{K}$ with the property 
  \begin{equation*}
  \int_{0}^{t}\, \vert\vert\, \phi(s)\, \vert\vert^2\, ds\, <\infty. 
  \end{equation*}
  Put $\phi_{t]}=\mathbbm{1}_{[0,t]}\, \phi$ and define in $\widetilde{\mathcal{H}}$ the operators 
  	\begin{eqnarray}
  	\label{4.4}
  	A_{\langle \phi\vert}(t)&=&a(\langle\,\phi_{t]}\vert) \\
  	\label{4.5}
  	A^\dag_{\vert\phi\rangle}(t)&=&a^\dag(\vert\,\phi_{t]}\rangle),\ \ \ \ 0\leq t<\infty.
  	\end{eqnarray}
  	where the right hand side vanish at $t=0$ and are respectively, the annihilation and creation fields evaluated at $\phi_{t]}.$ 
  	
  	Let now $t\rightarrow K(t)$ be a strongly measurable map from $\mathbb{R}_+$ into $\widetilde{\mathcal{H}}$ such that 
  	\begin{equation}
  	\label{4.6}
  \underset{0\leq s\leq t} {{\rm ess.\ sup.}}\ \vert\vert K(s)\vert\vert <\infty\ \ \forall \ t\geq 0.
  	\end{equation}
  Define the operators of multiplication $K_{s]}$ in $\widetilde{\mathcal{H}}$ by 
  \begin{equation*}
  \left(K_{s]}\, f\right) (t)=\, \mathbbm{1}_{[0,s]} (t)\, K(t)\, f(t),\ \ t\geq 0, \ f\in\mathcal{H}
  \end{equation*}
  and put 
  \begin{equation}
  \label{4.7} 
  \Lambda_K(t)=\lambda\left(K_{t]}\right),\ \ t\geq 0
  \end{equation}	
  where the right hand side is defined as the conservation field evaluated at $K_{t]}$. See equation (\ref{3.24}).
  
  The factorizability of the Weyl representation along with equations (\ref{3.10}), (\ref{3.11}) and (\ref{3.21}) show that, if we denote by $M(t)$, any one of the operators (\ref{4.4}), (\ref{4.5}), (\ref{4.7}), then the map $t\rightarrow M(t)\, e(f)$ is continuous for any $f$ in $\mathcal{H}$ and 
  \begin{equation} 
  \label{4.8}
  \left(M(t)-M(s)\right)\, e(f)=e(f_{s]})\otimes\, \left(M(t)-M(s)\right)\, 
  e(f_{[s,t]})\otimes e(f_{[t})
  \end{equation}
  for all $0<s<t<\infty$. In short, the map $t\rightarrow M(t)$ is a regular adapted process with property (\ref{4.8}) i.e., $\left(M(t)-M(s)\right)\,$ is effectively an operator in the component Fock space $\widetilde{\mathcal{H}}_{[s,t]}$. This suggests the interpretation that $\{M(t)\}$ is a {\em process with independent increments} in the continuous tensor product Hilbert space $\widetilde{\mathcal{H}}$.  
  
  We call $\{	A_{\langle \phi\vert}(t)\}$, 	$\{A^\dag_{\vert \phi\rangle}(t)\}$ and 	$\{\Lambda_{K}(t)\}$, the {\em annihilation, creation} and {\em conservation processes} with {\em strength} $\langle \phi\vert$, $\vert \phi\rangle$ and $K$ respectively. 
  
  Consider the Hilbert space $\mathbb{C}\oplus\mathcal{K}$ and express any bounded operator and its adjoint in it by its matrices 
  $$\left(\begin{array}{cc}
  \alpha & \langle u\vert  \\ 
  \vert v\rangle & K 
  \end{array}\right),\ \   
  \left(\begin{array}{cc}
  \bar{\alpha} & \langle v\vert  \\ 
  \vert u\rangle & K^\dag 
  \end{array}\right),   $$
 so that the operator acts on the vector $\vert \beta \oplus w\rangle $ to yield 
 $\left\vert \alpha\,\beta + \langle u\vert w\,\rangle  \oplus \beta\, v+
 K\, w\right\rangle$, with $\alpha,\, \beta$ being scalars and $u,\, v,\, w$ being elements from $\mathcal{K}$.  Let $N(\cdot )$ be a matrix valued function on $\mathbb{R}_+$ such that 
 \begin{equation}
 \label{4.9}
 N(t)=\left(\begin{array}{cc}
 \alpha(t) & \langle \phi(t)\vert  \\ 
 \vert \psi(t)\rangle & K(t) 
 \end{array}\right),   
 \end{equation}	 
where $\alpha(\cdot )$ is locally integrable, $\phi$ and $\psi$ are in $\mathcal{H}_{\rm loc}$ and $t\rightarrow K(t)$ is a Borel map from $\mathbb{R}_+$ into $\mathcal{B}(\mathcal{K})$ which is locally norm bounded so that (\ref{4.6}) holds. Denote by $N^\dag(\cdot)$ the adjoint matrix function so that 
$N^\dag(t)=\left(N(t)\right)^\dag$ for every $t$. Then $N^\dag(\cdot)$ also satisfies the same properties. We denote by $\mathcal{N}$ the space of all such matrix-valued functions and observe that $\mathcal{N}$ is a linear space, closed under the adjoint map `$\dag$'. 

For any $N(\cdot)$ in $\mathcal{N}$ given by (\ref{4.9})  define the adapted process $t\rightarrow \Lambda_N(t)$ by 
\begin{equation}
\label{4.10}
\Lambda_N(t)=\int_0^t\, \alpha(s)\, ds + A^\dag_{\vert \psi\rangle}(t)+\Lambda_K(t)+
A_{\langle \phi\vert}(t)  
\end{equation}  
where the last three summands on the right hand side are defined by (\ref{4.4}), (\ref{4.5}) and (\ref{4.6}). Then $\Lambda_N$ and $\Lambda_{N^\dag}$ are regular adapted processes, which are adjoint to each other on $\mathcal{E}(\mathcal{H})$ at every $t$ in $\mathbb{R}_+$. We use $\Lambda_N^\dag=\Lambda_{N^\dag}$ for any $N\in \mathcal{N}$. 

If $L$ is any adapted process in $\widetilde{\mathcal{H}}$, then $L(s)$ is effective in $\widetilde{\mathcal{H}}_{s]}$, $\Lambda_N(t)-\Lambda_N(s) $ is effective in $\widetilde{\mathcal{H}}_{[s,t]}$ and so, by (\ref{3.10}), (\ref{3.21}) and (\ref{4.8}), we have   
\begin{eqnarray}
\label{4.11}
&&\langle e(f)\vert L(s)\, \left(\Lambda_N(t)-\Lambda_N(s)\right)\, \vert e(g)\rangle = 
\langle e(f)\vert \left(\Lambda_N(t)-\Lambda_N(s)\right)\, L(s)\, \vert e(g)\rangle \nonumber \\ 
&&\ \ \ \ \  = \langle e(f_{s]})\vert L(s) \, \vert e(g_{s]})\rangle \, \langle e(f_{[s,t]})\vert\,  \left(\Lambda_N(t)-\Lambda_N(s)\right)\, \vert e(g_{[s,t]})\rangle \, \langle e(f_{[t})\vert\, e(g_{[t})\rangle \nonumber \\ 
&&\ \ \ \  =\int_s^t\, \left\{ \alpha(\tau) + \langle f(\tau)\vert \psi(\tau)\rangle + 
\langle f(\tau)\vert K(\tau)\vert g(\tau) \rangle + \langle \phi(\tau)\vert g(\tau)\rangle  \right\}\, \langle e(f)\vert L(\tau)\vert e(g)\rangle\,  d\tau . \nonumber \\
\end{eqnarray} 
Let now $L$ be a simple adapted process with respect to a partition $0=t_0<t_1<\ldots <t_n<\ldots $ where $t_n\rightarrow \infty$ as $n\rightarrow \infty$. Define 
\begin{equation}
\label{4.12} 
X(t)=\int_{0}^{t}\, L\, d\Lambda_N= \sum_{j=0}^{\infty}\, L(t_j)\,
\left(\Lambda_N\left(t_{j+1} \wedge t\right)-\Lambda_N\left(t_j\wedge t\right)\right), 
\end{equation}
where $a\wedge b={\rm min}\, (a, b)$ for any $a,b$ in $\mathbb{R}_+$. One may replace any partition by a finer one in the definition of $L$ and it will not change the sum on the right hand side of (\ref{4.12}) where all but a finite number of summands vanish for each $0<t<\infty .$ It now follows from (4.10) that 
\begin{equation}
\label{4.13}
\left\langle e(f)\left\vert \int_0^t\, L\, d\Lambda_N  \right\vert e(g)\right\rangle = 
\int_0^t\, \nu_N(f,g)(s)\, \langle e(f)\vert L(s)\, \vert e(g)\rangle\, ds 
\end{equation}
where 
\begin{equation}
\label{4.14}
 \nu_N(f,g)(s)=\alpha(s)\, + \langle f(s)\vert \psi(s)\rangle + 
 \, \langle f(s)\vert K(s) \vert g(s)\rangle + \langle \phi(s)\vert g(s)\rangle
\end{equation}
Furthermore, the map $t\rightarrow X(t)$ defined by (\ref{4.12}) is a regular adapted process. We call (\ref{4.13}) the {\em first fundamental formula}. 

Suppose $L'$ is another simple adapted process, $N'$ another element of $\mathcal{N}$, and 
\begin{equation}
\label{4.15} 
X'(t)=\int_{0}^{t}\, L'\, d\Lambda_{N'}, 
\end{equation}
Then a more laborious but simple computation shows that 
\begin{eqnarray}
\label{4.16}
&&\langle X(t)\, e(f)\vert X'(t)\, e(g)\rangle= \int_0^t \left\{ \langle L(s)\, e(f)\vert X'(s)\, e(g) \rangle [\nu_{N^\dag}(f,g)](s) \right.  \\ 
&& \ \ \ \ \ \ \left. +\langle X(s)\, e(f)\vert L'(s)\, e(g) \rangle [\nu_{N'}(f,g)](s)
  + \langle L(s)\,  e(f)\vert L'(s)\, e(g) \rangle [\nu_{N^\dag\circ N'}](f,g)](s) \right\}\, ds \nonumber 
\end{eqnarray}
for all $f,\ g$ in $\mathcal{H}$, where the operation $\circ$ in $\mathcal{N}$ is defined as follows. 

First, between matrices of the form  
 $$\left(\begin{array}{cc}
 \alpha_j & \langle u_j\vert  \\ 
 \vert v_j\rangle & K_j 
 \end{array}\right),\ \ j=1,2$$  
 where $\alpha_j$ is a scalar, $u_j,\ v_j\in\mathcal{K}$, $K_j\in \mathcal{B}(\mathcal{K}), \ j=1,2,$ define 
 \begin{eqnarray}
 \label{4.17}
 \left(\begin{array}{cc}
 \alpha_1 & \langle u_1\vert  \\ 
 \vert v_1\rangle & K_1 
 \end{array}\right)\ \circ \left(\begin{array}{cc}
 \alpha_2 & \langle u_2\vert  \\ 
 \vert v_2\rangle & K_2 
 \end{array}\right) &=& 
 \left(\begin{array}{cc}
 \alpha_1 & \langle u_1\vert  \\ 
 \vert v_1\rangle & K_1 
 \end{array}\right)\, \left(\begin{array}{cc}
 0 &  0  \\ 
 0 & I 
 \end{array}\right)  \,  \left(\begin{array}{cc}
 \alpha_2 & \langle u_2\vert \nonumber   \\ 
 \vert v_2\rangle & K_2 
 \end{array}\right) \\ 
 &=& \left(\begin{array}{cc}
 \langle u_1\vert v_2\rangle  & \langle u_1\vert\, K_2  \\ 
 K_1\, \vert v_2\rangle & K_1\,K_2 
 \end{array}\right)     
 \end{eqnarray}
 Now, for $N_1(\cdot)$, $N_2(\cdot)$ in $\mathcal{N}$, do the same operation $\circ$ pointwise at $t\in \mathbb{R}_+$. This determines $N_1\circ N_2$. With the operation $\circ$,  $\mathcal{N}$ becomes an associative algebra with involution $\dag$.  
 We call equation (\ref{4.16}) the {\em second fundamental formula}. 
 
 \noindent {\bf Definition 4.1}  An adapted process $t\rightarrow L(t)$ is said to be {\em stochastically integrable} if there exists a sequence of simple adapted processes $\{L_n,\ n=1,2,\ldots\}$ such that 
 \begin{equation}
 \label{4.18}
 \underset{n\rightarrow\infty}{\rm lim} \int_{0}^t \left\vert\left\vert\,  
 L_n(s)\, e(f) - L(s)\, e(f)\,
 \right\vert \right\vert^2\, \vert\vert\, g(s)\, \vert\vert^2\, ds=0
 \end{equation} 
 for every $f,\, g$ in $\mathcal{H}$. 
 
 If $L$ is stochastically integrable it is a consequence of the first and second fundamental formula that for any  approximating sequence $\{L_n\}$ of simple processes satisfying (\ref{4.18}) 
 \begin{equation*}
\underset{n\rightarrow\infty}{\rm lim}\, \left(\int_0^t\, L_n\, d\Lambda_N\,\right)\, e(f)
 \end{equation*} 
 exists for every $N\in\mathcal{N}$, $f\in\mathcal{H}$ and the limit is independent of the approximating sequence. We denote this limit by 
 \begin{equation*}
  \left(\int_0^t\, L\, d\Lambda_N\,\right)\, e(f),\ f\in \, \mathcal{H}.
 \end{equation*} 
 
 We denote by $\mathbb{L}(\mathcal{H})$, the linear space of all stochastically integrable processes. Then the following properties hold.  
 \begin{enumerate}[label=(\roman*)]
\item $X(t)=\int_0^t\, L\, d\Lambda_N,\ t\geq 0$ is a regular adapted process for each $L\in \mathbb{L}(\mathcal{H})$, $N\in\mathcal{N}$. Thus $X\in\mathbb{L}(\mathcal{H})$. 
\item The map $L\rightarrow \int_{0}^{t}\, L\, d\Lambda_N,$ for every $t\geq 0$ is linear in $\mathbb{L}(\mathcal{H})$. 
\item The map $N  \rightarrow \int_{0}^{t}\, L\, d\Lambda_N,$ for every $t\geq 0$ is linear in $\mathcal{N}$ for each  	$L\in \mathbb{L}(\mathcal{H})$.
 \item If $L$, $L^\dag \in \mathbb{L}(\mathcal{H})$ satisfy the condition that the operators  $L(t)$, $L^\dag(t)$ are adjoint to each other on the exponential domain $\mathcal{E}(\mathcal{H})$ for every $t$, then the processes 
\begin{eqnarray*}
	X(t)=\int_0^t\, L\, d\Lambda_N,\ \ X^\dag(t)=\int_0^t\, L^\dag\, d\Lambda_{N^\dag},\ \ t\geq 0
\end{eqnarray*}
are adjoint to each other in $\mathcal{E}(\mathcal{H})$ for every $t$. 
\item By the first fundamental formula for $L\in \mathbb{L}(\mathcal{H})$, $N\in \mathcal{N}$, 
\begin{eqnarray}
\label{4.19}
\frac{d}{dt}\,\left\langle e(f)\, \left\vert \int_0^t\, L\, d\Lambda_N\, \right\vert  e(g) \right\rangle & =& \nu_{N}(f,g)(t)\,  \langle e(f)\,\vert L(t)\,\vert e(g)\rangle, \ \ \ {\rm a.e.}(t)
\end{eqnarray}
where  $\nu_{N}(f,g)(t)$ is given by (\ref{4.9}) and (\ref{4.14}). 
\item If $L_j\in \mathbb{L}(\mathcal{H}),\ j=1,2$ and 
\begin{equation*}
X_j(t)=\int_0^t\, L_j\, d\Lambda_{N_j},\ j=1,2
\end{equation*} 
by the second fundamental formula 
\begin{eqnarray}
\label{4.20}
&&\frac{d}{dt}\,\left\langle X_1(t)\, e(f)\, \left\vert  \right.  X_2(t)\, e(g) \right\rangle = \left\langle L_1(t)\, e(f)\, \left\vert  \right.  X_2(t)\, e(g) \right\rangle\ \nu_{N_1^\dag}(f,g)(t) \nonumber \\ 
&&\hskip 0.5in \  \  +\langle X_1(t) e(f)\, \vert    L_2(t) e(g) \rangle\ \nu_{N_2}(f,g)(t) \nonumber \\
&&\hskip 0.5in \ +\langle L_1(t) e(f)\, \vert    L_2(t)\, e(g) \rangle\ \nu_{N_1^\dag\circ N_2}(f,g)(t),\ \   {\rm a.e.}(t)\ {\rm for\ all\ } f,g\in \mathcal{H}.
\end{eqnarray} 
\end{enumerate}
We now express an equation of the form 
\begin{eqnarray*}
X(t)=x_0\ I+ \int_0^t\, L\, d\Lambda_N,\ L\in\mathbb{L}(\mathcal{H}),\ N\in\mathcal{N},\ x_0\in\mathbb{C}	
\end{eqnarray*} 
in the differential notation, 
\begin{equation*}
dX=L\, d\Lambda_N,\ \ X(0)=x_0. 
\end{equation*}
 Then (\ref{4.19}) can be expressed as 
 \begin{equation*}
\langle e(f)\, \vert L\, d\Lambda_N\, \vert  e(g) \rangle
= \langle e(f)\, \vert L(t)\, \vert  e(g) \rangle\, \nu_N(f,g)(t)\, dt. 
 \end{equation*}
 Choosing $L(t)=I$ for all $t$, we have 
\begin{equation*}
\langle e(f)\, \vert  d\Lambda_N\, \vert  e(g) \rangle
= e^{\langle f\vert g\rangle}\,  \nu_N(f,g)(t)\, dt. 
\end{equation*}
If $L^\dag(t), \ t\geq 0$ is also in $\mathbb{L}(\mathcal{H})$, where $L^\dag(t)$ is adjoint to $L(t)$ in $\mathcal{E}(\mathcal{H})$ for every $t$, then 
\begin{equation*}
\left( L\, d\Lambda_N\,\right)^\dag= \left(  d\Lambda_N\,\right)^\dag\, L^\dag = \, 
L^\dag\,   d\Lambda_{N^\dag}.
\end{equation*} 
 Suppose, in property (vi), equation (\ref{4.20}), the operators $L_1(t)$ and $X_1(t)$ have adjoints with sufficiently large appropriate domains (including $\mathcal{E}(\mathcal{H})$) we can move them to the right within the brackets $\langle \ \ \rangle$ on both sides and rewrite it as, 
 \begin{eqnarray*}
 \langle e(f) \vert d(X_1^\dag(t)\, X_2(t)\vert e(g)\rangle &=& 
 \langle e(f) \vert L_1^\dag(t)\, d\Lambda_{N_1^\dag}\, X_2(t)\vert e(g)\rangle 
    +\langle e(f) \vert X_1^\dag(t)\, L_2(t)\,  d\Lambda_{N_2}\,\vert e(g)\rangle \\ 
    && \ \  + \langle e(f) \vert L_1^\dag(t)\, L_2(t)\,  d\Lambda_{N_1^\dag\circ N_2}\,\vert e(g)\rangle.  
 \end{eqnarray*}
Taking the liberty to drop $e(f),\ e(g),$ we express (\ref{4.20}) as 
\begin{equation}
\label{4.21}
d\,\left(X_1^\dag\, X_2\right)= \left(dX_1^\dag\right)\,  X_2+ X_1^\dag\, \left(dX_2\right)\,+ L_1^\dag\,L_2\, d\Lambda_{N_1^\dag\circ N_2}. 
\end{equation}
 We call the last summand on the right hand side the product differential $\left(dX_1^\dag\right)\, \left(dX_2\right)$ or simply $dX_1^\dag\, dX_2$. Of course, its rigorous interpretation is given by the second fundamental formula (\ref{4.20}), expressed only in the exponential domain.  
 
 In (\ref{4.21}) we drop the symbol $\dag$ and say that for processes $X_j$, satisfying, 
 \begin{equation*}
 dX_j=L_j\, d\Lambda_{N_j},\ \ L_j\in\mathbb{L}(\mathcal{H}), \ N_j\in\mathcal{N}, j=1,2, 
 \end{equation*} 
 the relation 
 \begin{eqnarray}
 \label{4.22}
 d\,X_1\, X_2&=& X_1\, dX_2+ (dX_1)\, X_2+ dX_1\, dX_2 \nonumber  \\ 
 &=& X_1\,L_2\,  d\Lambda_{N_2}+ L_1\,X_2\,  d\Lambda_{N_1}+ L_1\, L_2\, d\Lambda_{N_1}\, d\Lambda_{N_2} 
  \end{eqnarray}
 where $d\Lambda_{N_1}\, d\Lambda_{N_2}=d\Lambda_{N_1\circ N_2}$. 
 	
When $X_j=\Lambda_{N_j}$, equation (\ref{4.22}) holds in the rigourous sense. Otherwise, it is to be understood in the generalized sense of (\ref{4.20}). We call (\ref{4.22}) a {\em quantum It{\^o} formula}. The space 
$$\mathcal{I}(\mathcal{H})=\left\{\, d\Lambda_N,\, N\in\mathcal{N}\right\}$$ 
is an involutive and associative algebra, with the involution $\left(d\Lambda_{N}\right)^\dag=d\Lambda_{N^\dag}$ and product rule 
$d\Lambda_{N_1}\, d\Lambda_{N_2}= d\Lambda_{N_1\circ N_2},$ where the operation $\circ$ is as described in the discussion below equation (\ref{4.16}). We call $\mathcal{I}(\mathcal{H})$, the It{\^o} differential algebra in $\Gamma(\mathcal{H})$. 

\section{From Weyl representation to L{\'e}vy processes}

If $\mu,\, \nu$, are two probability measures and $\mu\times \nu$ is their product measure, there is a natural isomorphism between the Hilbert spaces $L^2(\mu\times \nu)$  and $L^2(\mu)\otimes L^2(\nu)$. If this simple idea is stretched to continuous tensor products, one can expect to realize the classical stochastic processes with independent increments through the continuous tensor product structure lodged in the time included boson Fock space. To construct stochastic processes in the quantum probabilitistic framework of Hilbert spaces and state vectors, one needs a meaningful collection of observables. Usually, they arise from a group representation. The boson Fock frame together with the Weyl representation of the euclidean group of the underlying Hilbert space and the machinery of quantum stochastic calculus is a gold mine for constructing stochastic processes. As a starting and also startling example, we shall illustrate how all the L{\'e}vy processes with independent increments arise from the factorizable Weyl representation along properly chosen paths in the Euclidean group of the Hilbert space over which the Fock space is erected.  
 
 We begin with a Hilbert space $\mathcal{H}$ of the form $\mathcal{H}=\mathcal{K}\otimes L^2(\mathbb{R}_+)$ and the associated boson Fock space $\widetilde{\mathcal{H}}=\Gamma(\mathcal{H})$ as in Section 4. Let $t\rightarrow (\phi(t),U(t))$ be a Borel map from $\mathbb{R}_+$ into the euclidean group $\mathbb{E}(\mathcal{K})$ such that $\phi\in\mathcal{H}_{\rm loc}$ i.e., $\phi$ is a locally norm square integrable $\mathcal{K}$-valued function on $\mathbb{R}_+$. For $0\leq a<b<\infty$, define 
   \begin{equation*}
   \phi_{[a,b]}(t)=\left\{ \begin{array}{ll} \phi(t) & {\rm if}\ t\in[a,b], \\ 
   	0 & {\rm otherwise},\end{array} \right.
   	\end{equation*}
   	 	and 
   	\begin{equation*}
   	\left(U_{[a,b]}\, f\right)(t)=\left\{ \begin{array}{ll} U(t)\, f(t) & {\rm if}\ t\in[a,b], \\ 
   		f(t) & {\rm otherwise}, \end{array} \right.
   		\end{equation*}
   		for every $f\in \mathcal{H}$. Write $\phi_{t]}=\phi_{[0,t]},$\ $U_{t]}=U_{[0,t])}, \ \forall \ t$.
   Define 
   \begin{equation}
   \label{5.1}
   W([a,b])=W\left(\phi_{[a,b]}, \, U_{[a,b]}\right),
   \end{equation}
   where the right hand side is the Weyl representation of $\mathbb{E}(\mathcal{H})$ at $\left(\phi_{[a,b]}, \, U_{[a,b]}\right)$. 
   
   \noindent {\em Lemma 5.1 :} For any $f,g$ in $\mathcal{H}$, 
   \begin{eqnarray} 
   \label{5.2}
   \langle e(f)\vert  W([a,b])\vert e(g) \rangle && = {\rm exp}
  \left\{ \int_a^b\left(-\frac{1}{2}\, \vert\vert\phi(s)\vert\vert^2 
   +\langle f(s)\vert \phi(s)\rangle \right)\, ds  \right.  \nonumber  \\ 
  && + \int_a^b\left( \langle f(s)\vert U(s)\vert g(s)\rangle 
  -\langle \phi(s)\vert U(s)\vert g(s)\rangle \right)\, ds \\
   &&  +\int_{0}^a \langle f(s)\vert g(s)\rangle\, ds + \left. \int_{b}^\infty \langle f(s)\vert g(s)\rangle\, ds  
   \right\},  \ \ \  0\leq a<b<\infty. \nonumber 
   \end{eqnarray} 
  \noindent {\em Proof:} This is straightforward from the definition of Weyl representation of Section~3. \ \ \ $\square$
  \medskip
  
  \noindent {\em Remark:} Equation (\ref{5.1}) and the nature of exponential vectors imply that $W([a,b])$ is effective only in the factor $\widetilde{\mathcal{H}}([a,b])$ of the tensor product factorization 
  	$$\widetilde{\mathcal{H}}= \widetilde{\mathcal{H}}([0,a]) \otimes \widetilde{\mathcal{H}}([a,b])\otimes \widetilde{\mathcal{H}}([b,\infty)).$$
  	If we write $f_{[a,b]}=\mathbbm{1}_{[a,b]}\, f$ and put 
  	\begin{equation*}
  	\gamma(a,b)=\langle e(f_{[a,b]}) \vert W(a,b)\vert e(g_{[a,b]})\rangle, 
  	\end{equation*}
  	then 
  \begin{equation*}
  \gamma(a,b)\, \gamma(b,c)=\gamma(a,c)\ \ {\rm for}\ a<b<c. 
  \end{equation*}
  This implies 
  $$W\left([a,b]\right)\, W\left([b,c]\right)=W\left([a,c]\right),\ 0\leq a<b<c<\infty.$$
  	
 \noindent {\em Theorem 5.2 :} The map $t\rightarrow W(t)=W\left([0,t]\right)$ defined by (\ref{5.1}) is a unitary, regular and adapted process obeying the quantum stochastic differential equation 
 \begin{equation}
 \label{5.3}
 dW=\, W\, d\Lambda_N,\ \ W(0)=1,
 \end{equation} 	
 where the matrix $N$ in $\mathcal{N}$ is given by  
 \begin{equation}
 \label{5.4} 
 N(t)=\left(\begin{array}{cc}
 -\frac{1}{2}\, \vert\vert \phi(t)\,\vert\vert^2 & -\langle \phi(t)\vert U(t) \\ 
 \vert \phi(t)\rangle & U(t)-1 
 \end{array}\right),\ \ \ t\geq 0. 
 \end{equation}
 
\noindent {\em Proof :} Putting $a=0,\ b=t$ in equation (\ref{5.2}) and differentiating with respect to $t$ (in the sense of absolute continuity) we get 
\begin{eqnarray*}
&& \, \frac{d}{dt}\, \langle e(f)\vert W(t)\vert e(g)\rangle = \langle e(f)\vert W(t)\vert e(g) \rangle \, \left\{-\frac{1}{2} \vert\vert \phi(t)\,\vert\vert^2 + \langle f(t)\vert \phi(t)\rangle + \langle f(t)\vert\left( U(t)-1\right) \vert g(t)\rangle \right. \\ 
&& \ \ \ \ \left.\frac{}{}  - 
\langle \phi(t)\vert U(t)\vert g(t)\rangle \right\}, \ \ {\rm a.e.}(t). 
\end{eqnarray*} 
This is nothing but the first fundamental formula of the equation 
\begin{equation*}
W(t)=1+\int_0^t\, W\, d\Lambda_N,
\end{equation*}
with $N(t)$ defined by (\ref{5.4}). See equations (\ref{4.19}), (\ref{4.9}), and (\ref{4.20}). Properties of adaptedness, regularity and unitarity are immediate. \ \ \ \ \ \ \ \ \hskip 1in  $\square$

\noindent {\em Remark :} We call the process $t\rightarrow W(t)$ in Theorem~5.2 the {\em Weyl process} at $(\phi, U)$  chosen in the begining. The {\em compensated} Weyl process 
\begin{equation*}
\widetilde{W}(t)=e^{\frac{1}{2}\, \int_0^t\, \vert\vert \phi(s)\vert\vert^2\, ds}\, W(t),\ \ t\geq 0
\end{equation*} 
satisfies the exponential {\em quantum martingale} equation 
\begin{equation*}
d\widetilde{W}=\widetilde{W}\, \left(dA^\dag_{\vert\phi\rangle} + d\Lambda_{U-1}-dA_{\langle \phi\vert\, U}\right),\ \ \widetilde{W}(0)=1. 
\end{equation*}
This is a perfect analogue of the classical exponential martingale 
\begin{equation*}
\xi(t)=e^{i\int_{0}^t\, \alpha\, dB+\frac{i}{2}\int_0^t\, \alpha^2\, dt}, 
\end{equation*}
  	satisfying the equation 
  	$$d\xi=i\, \alpha\, \xi\, dB $$ 
  	with $\alpha$ a real scalar valued function of time and $B$ the trajectory of a standard brownian motion process. 
  	
\noindent{\em Example 5.3 :} In Theorem~5.2, choose $U(t)=U_x(t)=e^{ix\, H(t)}$, where $H(t)$ is an observable in $\mathcal{K}$ such that the correspondence $t\rightarrow U_x(t)$ is a Borel map into $\mathcal{K}$, for every $x\in\mathbb{R}$. Choose an element $\psi$ in $\mathcal{H}_{\rm loc}$ and put 
\begin{equation*}
\phi_x(t)=U_x(t)\, \psi(t)-\psi(t),\ \ \ t\in \mathbb{R}_+. 
\end{equation*}
Then $\phi_x\in\mathcal{H}_{\rm loc}$. Now consider the Weyl process $\{W_x(t),\ t\geq 0\}$ at $(\phi_x,\, U_x)$. By the multiplication rule in the definition of Weyl representation, we have 
\begin{equation*}
W_x(t)\, W_y(t)=W_{x+y}(t)\, {\rm exp}\left( i \, {\rm Im}\, \int_0^t\, \langle 
\psi (s) \vert \left(U_{x+y}(s)-U_x(s)-U_y(s)\right)\vert \psi(s)\rangle ds \right). 
\end{equation*}  
Putting 
\begin{equation} 
\label{5.5}
\widetilde{W}_x(t)= W_x(t)\, {\rm exp}\left( i \, {\rm Im}\, \int_0^t\, \langle 
\psi (s) \vert U_{x}(s)\vert \psi(s)\rangle ds \right),  
\end{equation}   	
we observe that,   $\{\widetilde{W}_x(t),\ \ x\in\mathbb{R}\}$, for every fixed $t$, is a one parameter unitary group in $x$. Hence there exists an observable $Z(t)$ in $\widetilde{\mathcal{H}}$ such that 
\begin{equation} 
\label{5.6}
\widetilde{W}_x(t)= e^{i\, x\, Z(t)}.  
\end{equation}   	
From the factorizability of the Weyl representation, it follows that 
\begin{equation*}
 e^{i\, x\, \left(Z(t)-Z(s)\right)}=\widetilde{W}_x(t)\, \widetilde{W}_x(s)^\dag
\end{equation*}
and $Z(t)-Z(s)$ is an observable, which is effective only in the factor $\widetilde{\mathcal{H}}\left([s,t]\right)$ of the factorization 
\begin{equation*}
\widetilde{\mathcal{H}}=\widetilde{\mathcal{H}}\left([0,s]\right) \otimes 
\widetilde{\mathcal{H}}\left([s,t]\right) \otimes 
\widetilde{\mathcal{H}}\left([t,\infty)\right). 
\end{equation*} 
Furthermore 
\begin{equation*}
\langle e(0)\vert e^{ix\, \left(Z(t)-Z(s)\right)} \vert e(0)\rangle = 
{\rm exp}\left(\int_s^t\, \langle \psi(\tau) \vert \left(U_x(\tau)-1\right)\vert \psi(\tau)\rangle\, d\tau \right)
\end{equation*}
If $P^{(\tau)}$ denotes the spectral measure of $H(\tau)$, we can express the equation above as
\begin{equation*}
\langle e(0)\vert e^{ix\, \left(Z(t)-Z(s)\right)} \vert e(0)\rangle = 
{\rm exp}\left(\int_s^t\,\int_\mathbb{R}\,  (e^{ixy}-1)\, L(\tau, dy)\, d\tau \right)
\end{equation*}
where 
\begin{equation*}
 L(\tau, E)=\langle \psi(\tau)\vert P^{(\tau)}(E)\vert \psi(\tau)\rangle
\end{equation*}
for every Borel subset $E$ of $\mathbb{R}$. In other words, in the vacuum state $\vert e(0)\rangle$ the observable $Z(t)-Z(s)$ has the infinitely divisible distribution with L{\'e}vy measure $\int_s^t\, L(\tau, E)\, d\tau, \ E\subset\mathbb{R}$ varying in the Borel $\sigma$-algebra of $\mathbb{R}$. In short, $\{Z(t), t\geq 0\}$ executes, in the vacuum state, a L{\'e}vy process with independent increments, where the L{\'e}vy measure of $Z(t)$ is $\int_0^t\, L(\tau, \cdot)\, d\tau$. From Theorem~5.2 and equations (\ref{5.5})-(\ref{5.6}), we have the quantum stochastic differential equation 
\begin{eqnarray}
\label{5.7} 
d\, e^{ixZ}=e^{ixZ}\, d\Lambda_{N_x},\ \ Z(0)=0 
\end{eqnarray}
where 
 \begin{equation}
\label{5.8} 
N_x(t)=\left(\begin{array}{cc}
\langle \psi(t)\vert\left(e^{ixH(t)}-1\right)\vert \psi(t)\rangle & 
\langle \psi(t)\vert\left(e^{ixH(t)}-1\right) \\ 
\left(e^{ixH(t)}-1\right)\vert \psi(t)\rangle &  e^{ixH(t)}-1
\end{array}\right). 
\end{equation}
This also shows that 
\begin{equation*}
{\rm exp}\left(ixZ(t)-\int_0^t\,\langle \psi(\tau)\vert e^{ixH(\tau)}-1\vert \psi(\tau)\rangle\, d\tau \right),\ \ t\geq 0
\end{equation*}
is a {\em quantum} as well as classical martingale in the vacuum state. We call $\{Z(t)\}$ a L{\'e}vy process of type I. 

When $H(\cdot)$ is locally norm bounded differentiation with respect to $x$ in (\ref{5.7})-(\ref{5.8}) at $x=0$ shows that $Z(\cdot)$ has the form 
\begin{eqnarray*}
Z(t)=\int_{0}^t\, \langle \psi(\tau)\vert H(\tau)\vert \psi(\tau)\rangle\, d\tau + A^\dag_{H\vert \psi\rangle}(t)+\Lambda_H(t)+A_{\langle\psi\vert H}(t),\ \ t\geq 0. 
\end{eqnarray*}

\noindent {\em Example 5.4 :} Let $H(t),\, \psi(t)$ be as in Example~5.3, but 
\begin{equation*}
\phi_x(t)=\frac{e^{ixH(t)}-1}{H(t)}\, \psi(t),\ t\geq 0. 
\end{equation*}
Note that for any observable $B$, $\frac{e^{ixB}-1}{B}$, as a function of the 
selfadjoint operator $B$, is a bounded operator for any $x\in \mathbb{R}$ and $\phi_x$ as defined above is an element of $\mathcal{H}_{\rm loc}$. Now define the Weyl process $\{W_x(t), t\geq 0\}$ at $(\phi_x, U_x)$ using Theorem~5.2. Then the multiplication rule for Weyl representation shows that the map 
\begin{equation*}
x\rightarrow W_x(t)\, {\rm exp}\left( i\,\int_0^t\, \left\langle \psi(s) \left\vert  \frac{\sin\,x\, H(s)-x\, H(s)}{H(s)^2}\right\vert\psi(s)\right\rangle\, ds
\right)=W_x'(t)
\end{equation*}
yields a one parameter group of unitary operators in the parameter $x$. By the same arguments as in Example~5.3, we get an observable $Z'(t)$ such that 
\begin{equation}
\label{5.9}
e^{ix\,Z'(t)}=W_x'(t), \ \ \ t\in\mathbb{R}_+,\  x\in\mathbb{R},
\end{equation}
where $Z'(t)-Z'(s)$ is an observable effectively acting in the sector $\widetilde{\mathcal{H}}([s,t])$ of the Fock space $\widetilde{\mathcal{H}}$ for $0\leq s<t<\infty$. Furthermore, a simple computation with Weyl operators at $e(0)$ yields 
\begin{eqnarray}
\label{5.10} 
\langle e(0)\vert e^{ix\, \left(Z'(t)-Z'(s)\right)} \vert e(0)\rangle &=& 
{\rm exp}\left(\int_s^t\, 
 \left\langle \psi(\tau) \left\vert  \frac{e^{i\,x H(\tau)}-1-i\,x H(\tau)}{H(\tau)^2}\right\vert\psi(\tau)\right\rangle d\tau \right).
\end{eqnarray} 
If $P^{(\tau)}$ is the spectral measure of $H(\tau)$, (\ref{5.10}) can be rewritten as 
\begin{eqnarray}
\label{5.11} 
\langle e(0)\vert e^{ix\, \left(Z'(t)-Z'(s)\right)} \vert e(0)\rangle &=& 
{\rm exp}\left(\int_s^t\, \int_\mathbb{R} 
  \left(\frac{e^{i\,x y}-1-i\,x\,y}{y^2}\right)\, L(\tau, dy)\, d\tau \right).
\end{eqnarray} 
where 
\begin{equation}
\label{5.12}
L(\tau, E)=\langle \psi(\tau)\vert P^{(\tau)}(E)\vert \psi(\tau)\rangle,\ \forall\ \tau>0.
\end{equation}
In other words, in the vacuum state $\vert e(0)\rangle$, $\{Z'(t), t\geq 0\}$ executes a L{\'e}vy process, where the characteristic function of $Z'(t)$ is given by (\ref{5.11}) at $s=0$. 

Furthermore, Theorem~5.2 shows that 
\begin{equation*}
d\, e^{ixZ'(t)}=e^{ix\,Z'(t)}\, d\Lambda_{N_{x}'}(t),
\end{equation*}
 where the matrix $N_{x}'$ is given by 
\begin{equation*}
N_{x}'(t)=\left(\begin{array}{cc}
\left\langle \psi(t) \left\vert  \frac{e^{
i\,x H(t)}-1-i\,x H(t)}{H(t)^2}\right\vert\psi(t)\right\rangle & 
\left\langle \psi(t) \left\vert  \left(\frac{e^{i\,x H(t)}-1}{H(t)}\right)\right.\right. \\ 
\left.\left.\left(\frac{e^{i\,x H(t)}-1}{H(t)}\right)\right\vert \psi(t)\right\rangle &  e^{ixH(t)}-1
\end{array}\right). 
\end{equation*}
Thus, ${\rm exp}\left( ix\, Z'(t)-\int_0^t 
\left\langle \psi(\tau) \left\vert  \frac{e^{x H(\tau)}-1-i\,x H(\tau)}{H(\tau)^2}\right\vert\psi(\tau)\right\rangle\, d\tau\right)$ is a quantum as well as a classical martingale in the vacuum state. We say $\{Z'(t),\ t\geq 0\}$ is  a L{\'e}vy process of type II.

 When $H(\cdot)$ is locally norm bounded, differentiation with respect to $x$ at $x=0$ gives, 
$$Z'(t)=\left(\, A_{\vert \psi\rangle}^\dag +\Lambda_{H}+A_{\langle \psi\vert}\right) (t),\ \ \ \ \ t\geq 0. $$ 

\noindent{\em Remark :} If we choose Hilbert spaces $\mathcal{K}_j,\ j=1,2$, construct $\mathcal{H}_j,\ \widetilde{\mathcal{H}}_j$, choose $\psi_j$ in $\left( \mathcal{H}_j \right)_{\rm loc}$, maps $t\rightarrow H_j(t)$ from $\mathbb{R}_+$ into the space of observables in $\mathcal{K}_j$ according to Example~5.3, Example~5.4 and construct the type I L{\'e}vy process $Z_1(t)$ in $\Gamma(\mathcal{H}_1)$,  the type II L{\'e}vy process $Z_2'(t)$ in $\Gamma(\mathcal{H}_2)$ and pool them together as 
\begin{equation*}
Z(t)=Z_1(t)\otimes \mathbbm{1}_{\widetilde{\mathcal{H}}_2}
+ \mathbbm{1}_{\widetilde{\mathcal{H}}_1}\otimes Z'_2(t) 
\end{equation*}  
in  $\Gamma(\mathcal{H}_1)\otimes \Gamma(\mathcal{H}_2)=\Gamma(\mathcal{H}_1\oplus \mathcal{H}_2)$ where $\mathcal{H}_1\oplus \mathcal{H}_2=(\mathcal{K}_1\oplus \mathcal{K}_2)\otimes L^2(\mathbb{R}_+)$ we get a L{\'e}vy process in 
$\Gamma(\mathcal{H}_1\oplus \mathcal{H}_2)=\widetilde{\mathcal{H}_1\oplus \mathcal{H}_2}.$ By such combination, every L{\'e}vy process is achieved {\em modulo} a non-random drift. 

\section*{Acknowledgements} 

I am deeply indebted to V.~S.~Varadarajan for his masterly exposition of the Weyl-Wigner-Mackey philosophy of symmetry and group representations in quantum theory in his lectures at the Indian Statistical Institute, Calcutta (now Kolkata) during the years 1963-65. The notes of these lectures are embedded in his book {\em Geometry of quantum theory}~\cite{Va}. I thank Ray Streater for introducing me to the interplay between infinitely divisible probability distributions, continuous  tensor products of HIlbert spaces and group representations through several conversations on the papers~\cite{Ar}, \cite{AW}, and \cite{St}. I am happy to have collaborated with Klaus Schmidt on the topic of Fock space and probability, which resulted in~\cite{PS}. I thank Martin Lindsay for our collaboration in \cite{LiP} which resulted in the new clothing for the presentation of quantum It{\^o}'s formula. Finally a Himalayan thanks to Robin Hudson for a collaboration in the development of quantum stochastic calculus~\cite{HuP, KRP}, extending for more than two decades. Its influence can be felt very much in the whole story of this article. 

I am grateful to A.~R.~Usha Devi for her continued enthusiasm to learn the subject of quantum stochastic calculus and her willingness to typeset my handwritten manuscript in LaTeX.    

\bibliography{LevyProcesses_KRP_arXiv}

\end{document}